\documentclass[10pt, a4paper, conference]{IEEEtran}

\usepackage[utf8]{inputenc}
\usepackage[english]{babel}
\usepackage[nolist]{acronym}
\usepackage{mdframed}
\usepackage{newfloat}
\usepackage{tabularx}

\usepackage{subfigure}
\usepackage{courier}
\usepackage{epsfig}
\usepackage[justification=RaggedRight,singlelinecheck=false]{caption}
\usepackage{microtype}
\usepackage{pifont}
\usepackage{caption}
\usepackage{flushend}

\let\OLDthebibliography\thebibliography
\renewcommand\thebibliography[1]{
  \OLDthebibliography{#1}
  \setlength{\parskip}{0pt}
  \setlength{\itemsep}{2pt plus 0.3ex}
}

\newcommand{\eg}{{e.g., }}

\newcommand{\blackone}{\ding{182}}
\newcommand{\blacktwo}{\ding{183}}
\newcommand{\blackthree}{\ding{184}}
\newcommand{\blackfour}{\ding{185}}
\newcommand{\blackfive}{\ding{186}}

\begin{document}

\begin{acronym}
\acro{as}{Acceptable Source}
\acro{DHT}{Distributed Hash Table}
\acro{DNS}{Domain Name System}
\acro{DHT}{Distributed Hash Table}
\acro{CNRI}{Corporation for National Research Initiatives}
\acro{DOI}{Digital Object Identifier}
\acro{DONA}{Data Oriented Network Architecture}
\acro{EPIC}{European Persistent Identifier Consortium}
\acro{FTP}{File Transfer Protocol}
\acro{GHR}{Global Handle Registry}
\acro{GWDG}{Gesellschaft f\"ur wissenschaftliche Datenverarbeitung G\"ottingen mbH}
\acro{HTTP}{Hypertext Transport Protocol}
\acro{IETF}{Internet Engineering Task Force}
\acro{ICN}{Information Centric Networks}
\acro{IoT}{Internet of Things}
\acro{JSON}{Java Script Object Notation}
\acro{LDDA}{Location Dependent Data Access}
\acro{LHS}{Local Handle System}
\acro{PKI}{Public Key Infrastructure}
\acro{NETINF}{Network of Information}
\acro{NDN}{Named Data Networking}
\acro{NLSR}{Named-data Link State Routing}
\acro{PEX}{Peer Exchange}
\acro{PEM}{Privacy Enhanced Mail}
\acro{PID}{Persistent Identifier}
\acro{PIM}{Personal Information Manager}
\acro{PURSUIT}{Publish-Subscribe Architecture}
\acro{REST}{Representational State Transfer}
\acro{TCP}{Transmission Control Protocol}
\acro{UDP}{User Datagram Protocol}
\acro{URI}{Uniform Resource Identifier}
\acro{URL}{Uniform Resource Locator}
\acro{URN}{Uniform Resource Name}
\acro{WORM}{Write Once Read Multiple}
\acro{xt}{Exact Topic}
\end{acronym}

\title{On Constructing Persistent Identifiers \\with Persistent Resolution Targets}

\author{\IEEEauthorblockN{Oliver Wannenwetsch}
\IEEEauthorblockA{Gesellschaft f\"ur wissenschaftliche\\Datenverarbeitung G\"ottingen (GWDG),\\G\"ottingen, Germany\\
oliver.wannenwetsch@gwdg.de}
\and
\IEEEauthorblockN{Tim A. Majchrzak}
\IEEEauthorblockA{Department of Information Systems\\University of Agder,\\Kristiansand, Norway\\
timam@uia.no}
}

\maketitle

\begin{abstract}
Persistent Identifiers (PID) are the foundation referencing digital assets in scientific publications, books, and digital repositories.
In its realization, PIDs contain metadata and resolving targets in form of URLs that point to data sets located on the network.
In contrast to PIDs, the target URLs are typically changing over time;
thus, PIDs need continuous maintenance -- an effort that is increasing tremendously with the advancement of e-Science and the advent of the Internet-of-Things (IoT).
Nowadays, billions of sensors and data sets are subject of PID assignment.
This paper presents a new approach of embedding location independent targets into PIDs that allows the creation of maintenance-free PIDs using content-centric network technology and overlay networks.
For proving the validity of the presented approach, the Handle PID System is used in conjunction with Magnet Link access information encoding, state-of-the-art decentralized data distribution with BitTorrent, and Named Data Networking (NDN) as location-independent data access technology for networks.
Contrasting existing approaches, no green-field implementation of PID or major modifications of the Handle System is required to enable location-independent data dissemination with maintenance-free PIDs.
\end{abstract}


\IEEEpeerreviewmaketitle

\section{Introduction}
\label{sec:Introduction}

The concept of \textit{\ac{PID}} is essential for referencing, citing and linking (digital) resources using a durable and reliable identifier.
PIDs are used to ensure the \mbox{long-term} valid access to possibly moving digital resources that suffer from changing \acs{URL}s and storage locations in networks.
\acs{PID}s contain an adjustable target \ac{URL}.
To reflect changing and volatile data locations, the target \ac{URL} of a \ac{PID} must be updated to the currently valid storage location.
By employing organizational and technical measurements, different PID systems allow building, using and maintaining a \mbox{long-term} existing (digital) identifier that is backed by distributed systems, replication schemes, and policies.
With these measurements in place, \ac{PID} infrastructure operating organizations are able to offer \ac{PID} systems that are resilient against failure, and even catastrophic scenarios.
Ideally, the range of \ac{PID} infrastructure resilience includes scheduled downtimes of server and networks, major infrastructure problems caused by power failures, and hardware and software problems, as well as ultra critical events of complete data center losses caused by fires, explosions or natural disasters.

\begin{figure}[ht]
	\centering
	\includegraphics[width=0.48\textwidth]{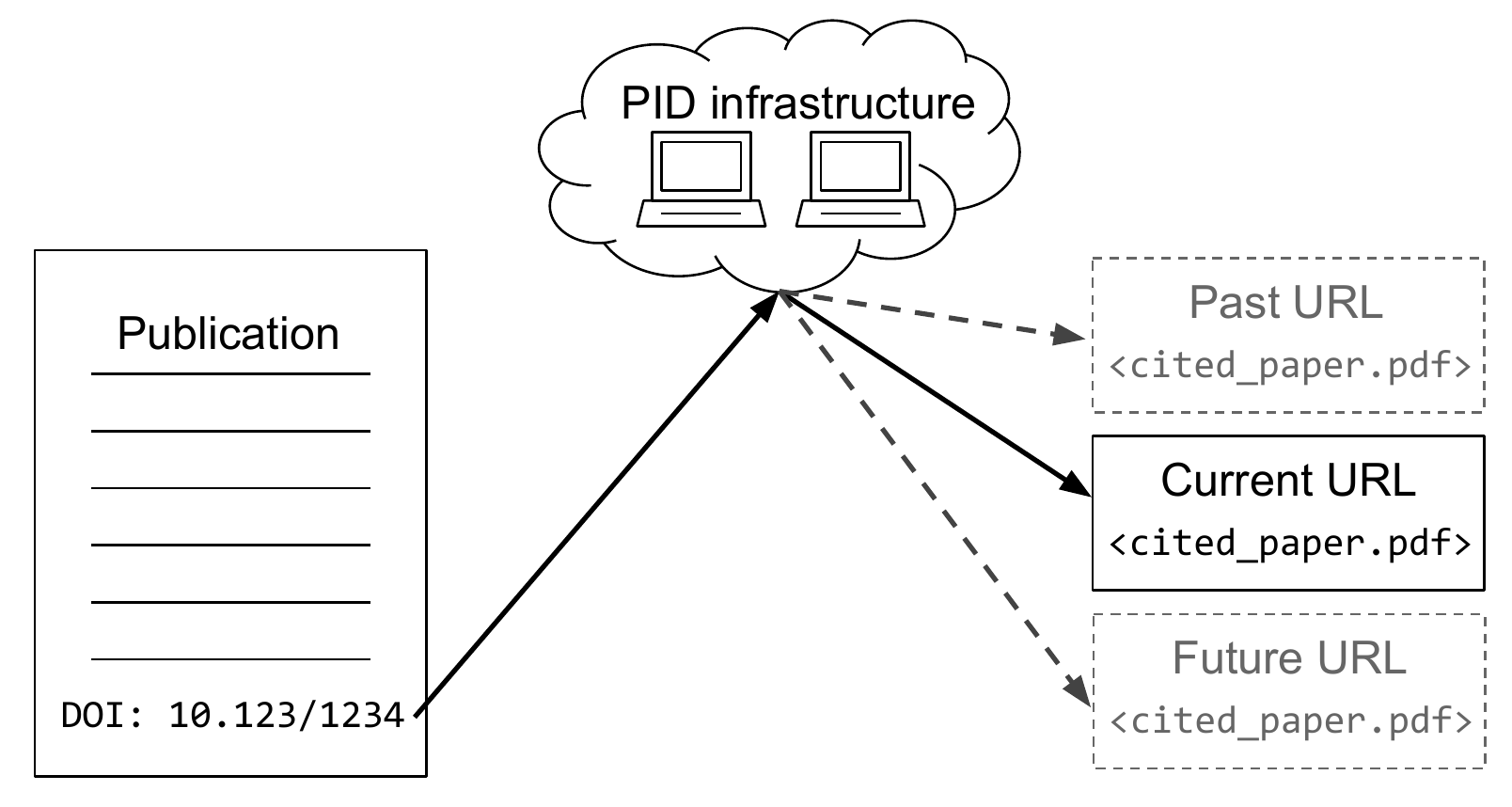}
	\caption{Adjustment of \ac{PID} target URLs to reflect current data location.}
	\label{fig:PIDPrinciple}
\end{figure}

Besides the measurements that protect \acs{PID}s on infrastructure level, \acs{PID}s have to be protected on the content-side as well.
The content of the \ac{PID} has to be intact and readable with given encoding schemes.
Furthermore, the metadata have to be addressed with a given metadata scheme that explains the semantics of the metadata.
Then, the content has to reflect the current state of the data object the \ac{PID} is linking to.
This includes the \emph{up-to-dateness} of metadata sets stored in \acs{PID}s, which are often encoded as \emph{key-value} pairs.
Particularly important is the correctness of the \ac{PID} metadata field \texttt{target URL}, which points to the digital object addressed by the \ac{PID} for long-term access (c.f. \mbox{Figure~\ref{fig:PIDPrinciple}}).
Contrasting the infrastructure protection of \ac{PID}, this content validation is not done by the \ac{PID} infrastructure operating organization.
It is a task of the data owners that registered the \ac{PID} for their data or the subsequent organization that has the task of curating the data and its associated \ac{PID}s.
Only those organizations have the necessary understanding on the data and its metadata to check and update \ac{PID}s for assuring long-term access.
Moreover, they are aware of the current location of the data linked by \ac{PID}.
Thus, with regular control and adjustment of the target URLs to the current data location, well-established PID systems can guarantee persistency of identifiers physically, while data owner organizations have to accept the burden of regularly checking and updating metadata and target URLs.
Only with collaboration \acs{PID}s are able to provide long-term access \cite{paskin_digital_2011}.

While \ac{PID}s solve the problem of changing data locations by constant efforts from \ac{PID} infrastructure providers and data hosting organizations, the network research community has come up with numerous concepts for creating location-independent data access.
In these concepts that access information for data attached to a network is not based on the data location, but it is based on the content of the data \cite{ahlgren_survey_2012}.
Hence, in the case of changing data location the access information remain stable.
Two efforts that realize location-independent access is the state-of-the-art decentralized data distribution technology \textit{BitTorrent} and \textit{Named Data Networking (NDN)} as next-generation Internet technology.
Although both technologies allow stable location-independent access to data, they do not provide persistent access like \ac{PID}.

Our approach presented in this paper combines the very stable concept of the \emph{Handle} \ac{PID} system with the advantages of location-independent access with its stable access information scheme.
By this, we significantly lower the efforts of \ac{PID} maintenance, as regular checks for data hosting organizations concerning locations and data availability will become obsolete.
\acs{PID}s that are enabled for location-independent access remain valid as long as data is available online in BitTorrent or \ac{NDN} networks.
To store the location-independent access information in existing Handle \acs{PID} data structures, we extend the -- yet not standardized -- approach of \emph{Magnet Link Schemes}, which is very successful in peer-to-peer communities \cite{van_der_sar_pirate_2009}.
In detail, our novel approach has the benefits of using an unmodified version of the Handle \ac{PID} system that allows a practical implementation of our approach in existing Handle \ac{PID} systems.
Besides the advantages of interoperability, it does not come with a significant impact on \ac{PID} resolution.
For the location-independent access with BitTorrent and \ac{NDN} only the small overhead of \ac{PID} resolution is added.

Furthermore, we extend the Magnet Link scheme into the domain of \ac{NDN} and provide a transport container format that includes besides the \ac{NDN} data name also the necessary cryptographic access information for verifying data access authenticity.
Thus, we can also improve Web browser-based \ac{NDN} applications that use HTML-links for interconnecting \ac{NDN} Web resources by the Magnet Link scheme. The latter has been originally designed for interconnecting non HTTP-resources in hypertext document contexts.

This paper is structured as follows.
First, we present existing efforts in Section~\ref{sec:RelatedWork}.
Second, we summarize the existing approaches of location-based data access through \ac{PID} in Section~\ref{sec:LocationBasedPIDTargets}.
Then, we line out state-of-the-art techniques for location-independent access in Section~\ref{sec:LocationIndependetAccess}.
In Section~\ref{sec:MagnetURIScheme}, we introduce the Magnet \ac{URI} scheme format as container for storing location-independent access information.
After that, we extend the Magnet \ac{URI} scheme for the usage in the domain of content-centric networks and \ac{PID} in Section~\ref{sec:LocationIndependentDataAccessThroughPID}.
A proof-of-concept implementation is illustrated in Section~\ref{sec:Implementation} together with an evaluation of performance in Section~\ref{sec:EvaluationAndDiscussion} that is combined with a discussion of the results.
Finally, we draw a conclusion in Section~\ref{sec:Conclusion}.

\section{Related Work}
\label{sec:RelatedWork}
To our knowledge, the concept of building care-free persistent identifier targets using content-centric technology has not been subject of extensive study.
In the literature several related concepts can be identified.

The concept of bridging different content-centric network systems through a centralized \ac{URN} system has been initially drafted by Sollins in 2012 \cite{sollins_pervasive_2012}.
Her concept utilizes foundations of \ac{PID} principles for creating an identification system for different \ac{ICN} families and their related data objects that meets the requirements of scalability, longevity, evolvability, and security.
Sollins's identification system abstracts different object naming schemes from \ac{ICN} families such as \ac{DONA} \cite{koponen_data-oriented_2007}, \mbox{\ac{NETINF} \cite{dannewitz_opennetinf_2012}} and \mbox{\ac{PURSUIT} \cite{fotiou_illustrating_2012}.}
Although, \ac{PID} principles are used for location-independent data access, the publication by Sollins does not suggest access through an existing well-introduced \ac{PID} system that is provided in our work, but rather uses a greed-field approach for location-independent data access.

The realization of complex secure naming schemes for content-centric data has been covered by Dannewitz et al. in 2010 \cite{dannewitz_secure_2010}.
They demand name persistency without incorporating the concept of \acs{PID}s.
In their publication, Dannewitz et al. clarify that basic security functionality must be attached directly to the data and its naming scheme, because the identity of network locations cannot be used as a trust base for data authenticity.
Our approach follows this principle for secure location-independent data access and facilitates directly attached \ac{PID} security mechanisms.
By this, location-independent access through \ac{PID} is shifted into the requirements formulated by Dannewitz et al., and our approach enables authentic data access through \acs{PID}s.

In the context of semantic digital archives for archiving data of \ac{PIM} applications, Haun and Nürnberger proposed a \ac{PID} schema for accessing objects in file systems using an URN-like Magnet Links \mbox{scheme \cite{haun_towards_2013}.}
They link the congruent attributes of the Magnet Link scheme to the attributes provided by some \ac{PID} systems such as global uniqueness, persistence and scalability for the application in offline data archives serving data from archive medium such as file systems on \ac{WORM} medium.
In contrast, our approach relies on currently employed \ac{PID} systems and incorporates location-independent data access in a distributed online environment using the full-featured Handle \ac{PID} system.

\section{Location-based PID Targets}
\label{sec:LocationBasedPIDTargets}

Today's data dissemination is dominated by end-to-end connections, URLs, and DNS-backed domain names.
When data is moved from one host to the other, it results in broken URLs and inaccessible content.
To ease these problems, \ac{PID}s are used for long-term data access by providing a long-living identifier.
When using a \ac{PID}, the identifier is embedded into a medium such as scientific publications, books, or Web sites.
To access the digital resources, \textit{behind} the \ac{PID} a resolution service is employed that uses the \ac{PID} to provide a currently valid network location (target URL).
\mbox{Figure~\ref{fig:PIDPrinciple}} illustrates how the target URL is adjusted to the current location when the data behind the \ac{PID} is moved from one host to the other.
Thus, \acs{PID}s reflect the current location of data and the identifier on the medium can remain unchanged.

\begin{figure}[ht]
\centering
\includegraphics[width=0.3\textwidth]{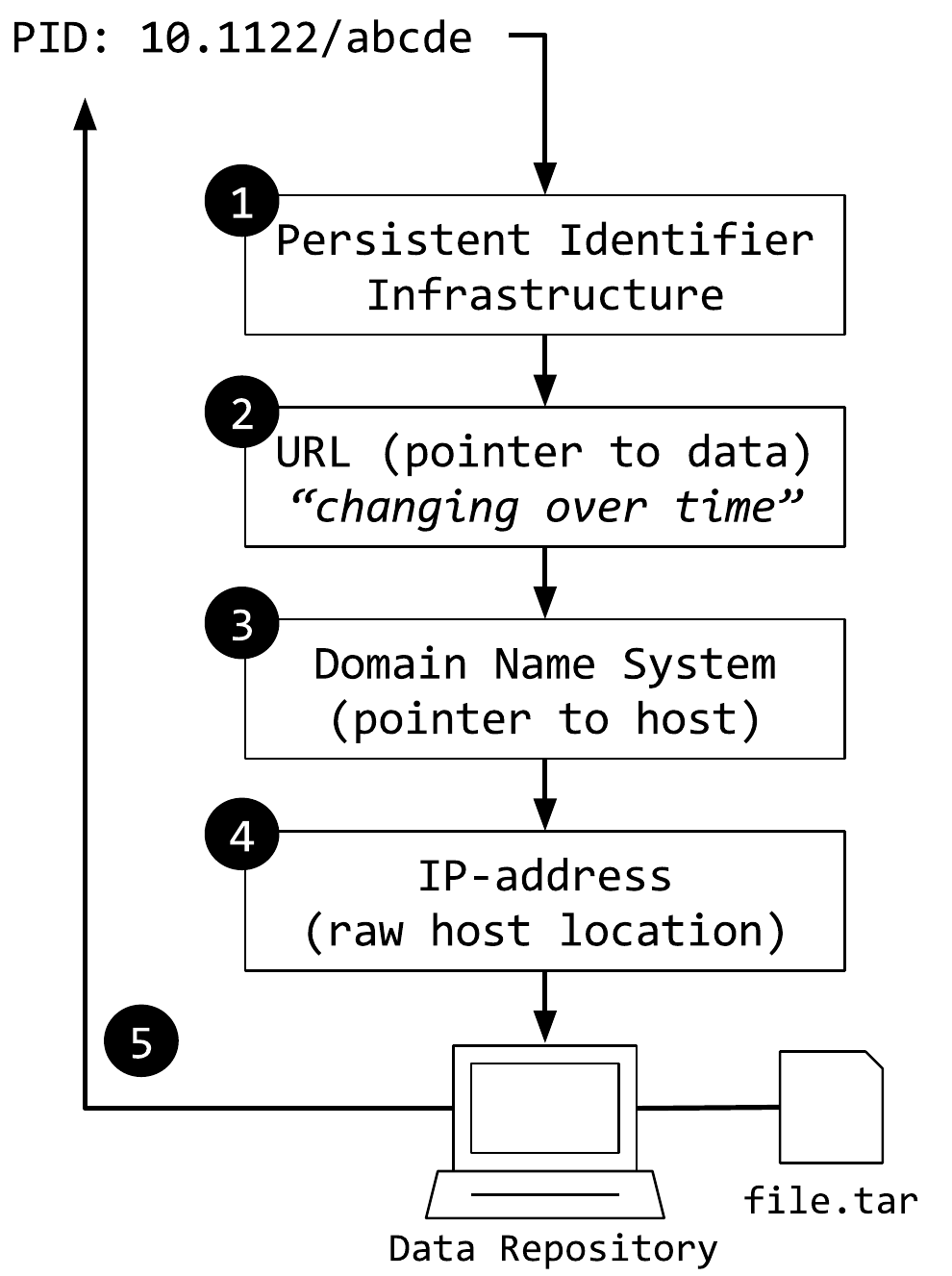}
\caption{Data access through PID requires a working chain of services relying on unstable URLs.}
\label{fig:PIDMaintenance}
\end{figure}

The location-dependent data access through target URLs stored in PIDs also forms the chain of PID resolution.
For this we have a look at the resolution chain presented in \mbox{Figure~\ref{fig:PIDMaintenance}}.
It depicts the fact that data access through \ac{PID} with \ac{HTTP} relies on different infrastructures and involves five levels from the \ac{PID} resolution up to the data download.
Even if the chain is shortened, e.g. by directly linking PID targets to IP-addresses instead of DNS-based host names, the problem remains identical:
If \ac{PID}-tagged data is moved from one host to the other, the \ac{PID} access chain needs to be adjusted on one or even on multiple levels to provide valid \ac{PID} resolution.
If the adjustment is not done or partially incorrect, location-depended data access is impossible through \ac{PID}.
Defects can occur on every level.
On \mbox{level \blackone~the} \ac{PID} HTTP resolution service can be temporary out of order.
The target \ac{URL} does not reflect the current location of the data in \mbox{stage~\blacktwo.}
On \mbox{level \blackthree~the} \ac{DNS} resolution can fail if a domain has been expired or DNS resolution fails due to misconfiguration.
Level \blackfour~and \blackfive~are related to the network, but their functionality is also required for successful data access through \ac{PID}.
To detect broken \ac{PID} resolution a check of every \ac{PID} target and, thus, a successful resolution is necessary for judging the integrity.
The check is done in many cases by evaluating the \ac{HTTP} status codes like \texttt{404}~\mbox{\texttt{-~not found}} that are provided by the data repositories \mbox{software \cite{hilse_implementing_2006}.}
This can only be achieved by regularly checking \emph{all} \acs{PID}s of an organization. This is very time consuming since typically robot or spider programs crawl all \ac{PID}s of a data owner.
The crawling programs are programmed and operated by repository owners and not \ac{PID} infrastructure providers. They provide  an optional data quality assurance service.

The adjustment of target URLs has impact on different dimensions and is shared unevenly between the users -- the \ac{PID} operators and data repository owners.
The costs and efforts behind \ac{URL} adjustments have been accepted as part of the \ac{PID} operation.
They are considered \emph{inevitable}, such as energy leaks in today's electrical grid infrastructure.
The adjustment of target URLs is a shared effort on the side of data owners and dependent on the number of \acs{PID}s a data owner has registered.

\begin{figure}[ht]
\centering
\includegraphics[width=0.49\textwidth]{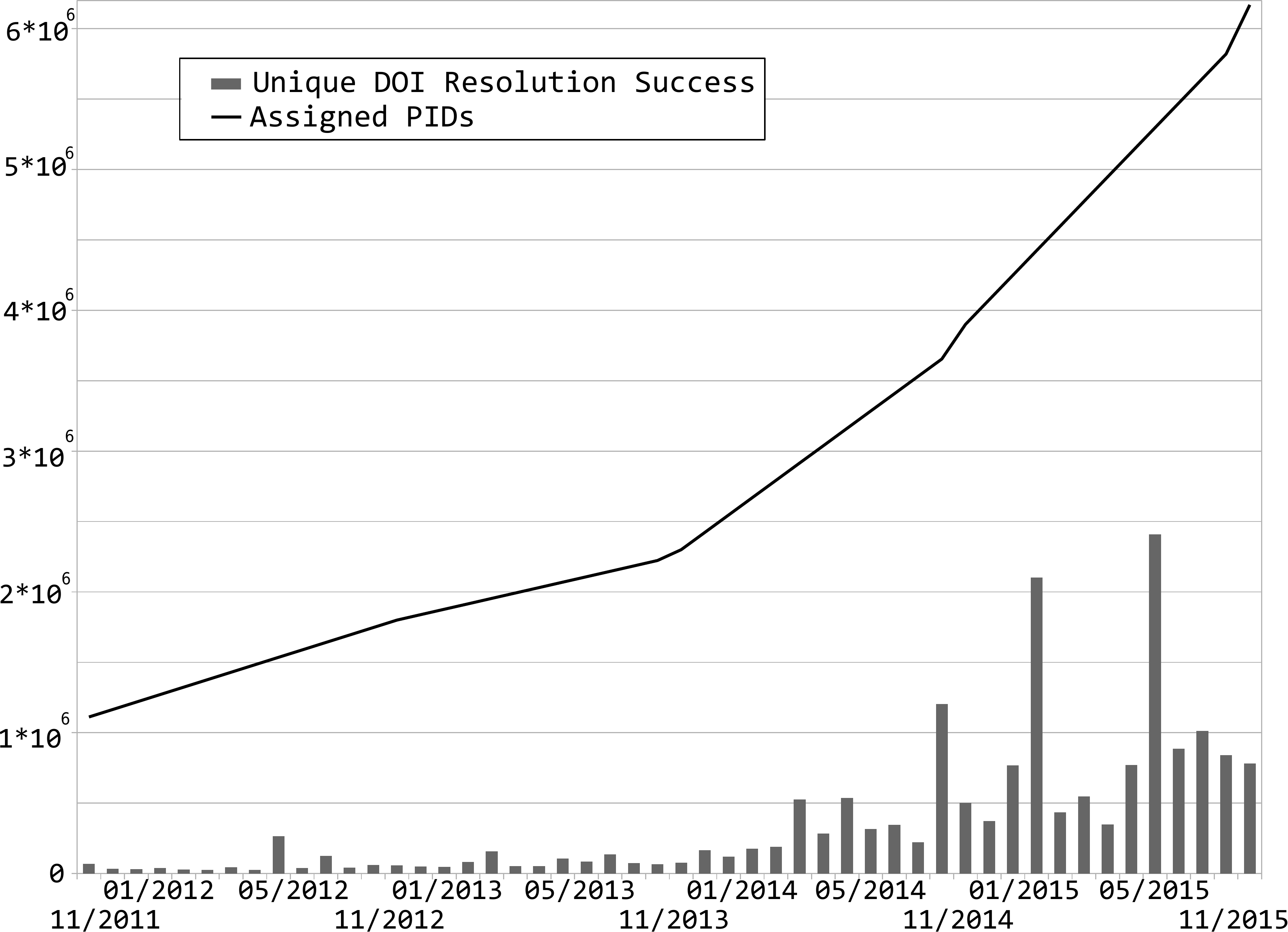}
\caption{PID assignment and unique successful resolution for the DataCite DOI infrastructure between 11/2011 and 11/2015 based on data from \cite{cruse_general_2016} \cite{_datacite_2016}.}
\label{fig:DOINumbers}
\end{figure}

It can be questioned whether the proliferation of e-Science already increases the effort necessary for \ac{PID} maintenance.
We thus have visualized statistics from DataCite (one of the largest PID infrastructure providers) in \mbox{Figure~\ref{fig:DOINumbers}.}
The assignment of new \acs{PID}s (\emph{line}) massively increased, following a super-linear pattern \cite{cruse_general_2016}.
An aggregation of the DataCite Statistics for successful DOI PID resolutions shows also a massive increase in PID-tagged data sets (\emph{bar}) \cite{_datacite_2016} \cite{fenner_digging_2015}.
With a massive increase of \ac{PID} numbers, the efforts for maintaining \ac{PID} targets will increase identically, as every assigned \ac{PID} needs to be checked for validity to comply with \ac{PID} infrastructure policy.
It is not a question, whether the \ac{PID} systems are scaling out sufficiently, but rather the data hosters are able to verify their location-dependent \ac{PID}s with a reasonable effort regularly.
PIDs with location-independent targets decouple the growing number of \acs{PID}s from the efforts of maintaining \ac{PID} target URLs.

\section{Location-Independent Access}
\label{sec:LocationIndependetAccess}

For location-independent data access, we propose two different techniques that allow access based on the content and not on the network location. We can thus show that our approach of location-independent persistent \ac{PID} resolution targets works with various location-independent access technologies.

\begin{figure}[t]
\centering
\includegraphics[width=0.49\textwidth]{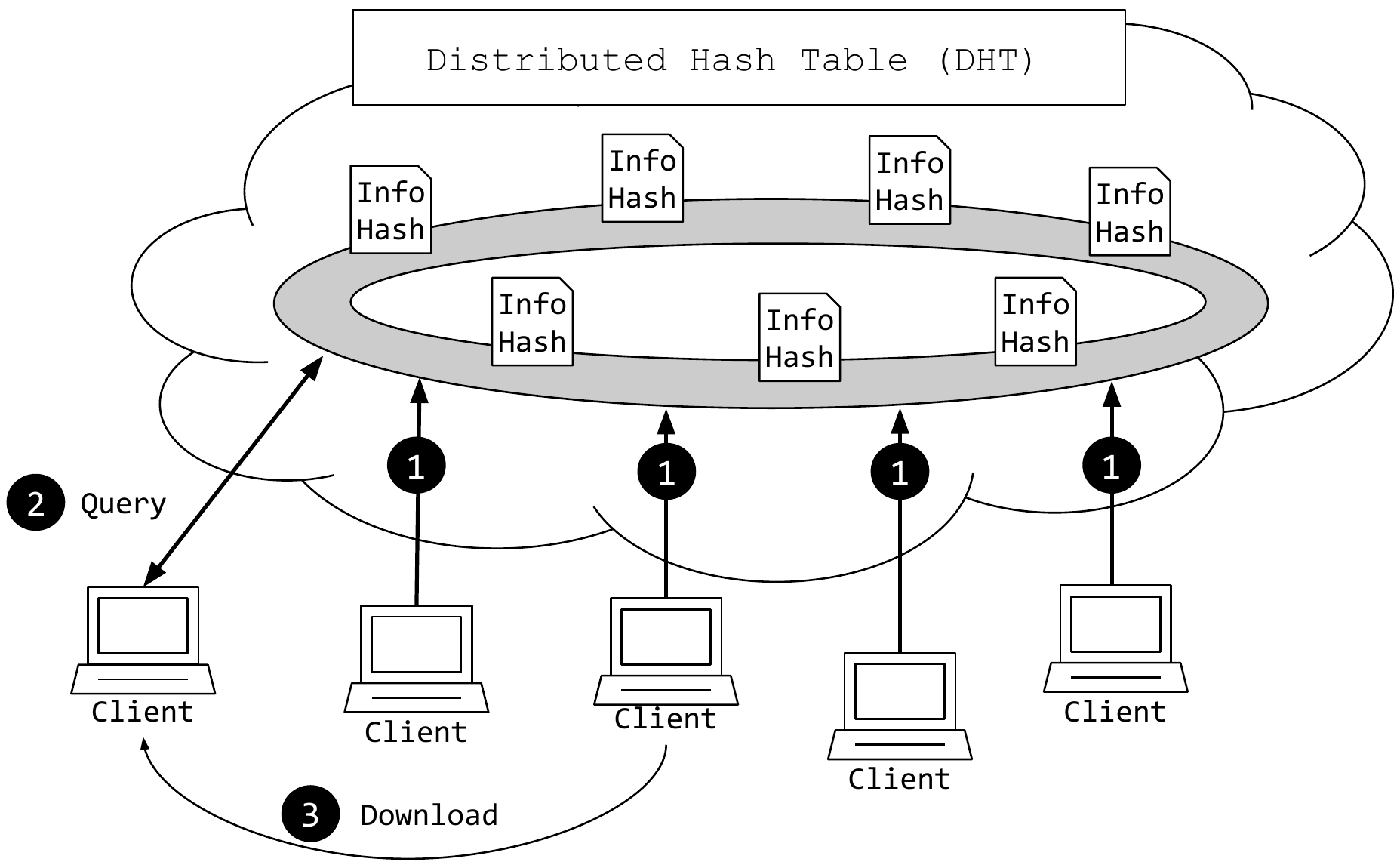}
\caption{Accessing data from a \ac{DHT}-controlled swarm using an infohash in BitTorrent.}
\label{fig:DHTSwarm}
\end{figure}

We propose BitTorrent, a well-established location-independent access technology.
It works on top of today's location-based networks with \ac{TCP} and \ac{UDP} \cite{cohen_bittorrent_2013}.
In contrast to existing location-based data repositories that are subject of PID target resolution, BitTorrent technology uses a peer-to-peer approach supporting parallel downloads.
With its latest features of \ac{DHT} and \ac{PEX}, BitTorrent does not require central infrastructure to discover other network peers and localize files \cite{loewenstern_bittorrent_2013} \cite{maymounkov_kademlia:_2002}.
Thus, data can be accessed with BitTorrent from every connected peer, as long as data is available online.
For addressing data sets, BitTorrent uses \emph{infohashes} that are computed as SHA-1 checksums on the content of the file.
Every peer that possesses the infohash can download the data set from the BitTorrent \emph{swarm} that consists of the peers offering the data set for download.
The swarm arrangement and the \emph{overlay network} for the specific file is computed for every download.
In \mbox{Figure~\ref{fig:DHTSwarm},} the BitTorrent file access is depicted.
In \mbox{step \blackone,} every node is sending its network address and the infohashes of the files ready for upload to the \ac{DHT}.
Then, a client can look up the infohash in the \ac{DHT} \mbox{(step \blacktwo)} to locate other peers that are able to serve the file that belongs to the infohash (or at least parts of the file).
Through connecting to the peers in step \blackthree, the client is retrieving the file.
This can be done simultaneously by parallel peer connection.

\begin{figure}[t]
\centering
\includegraphics[width=0.49\textwidth]{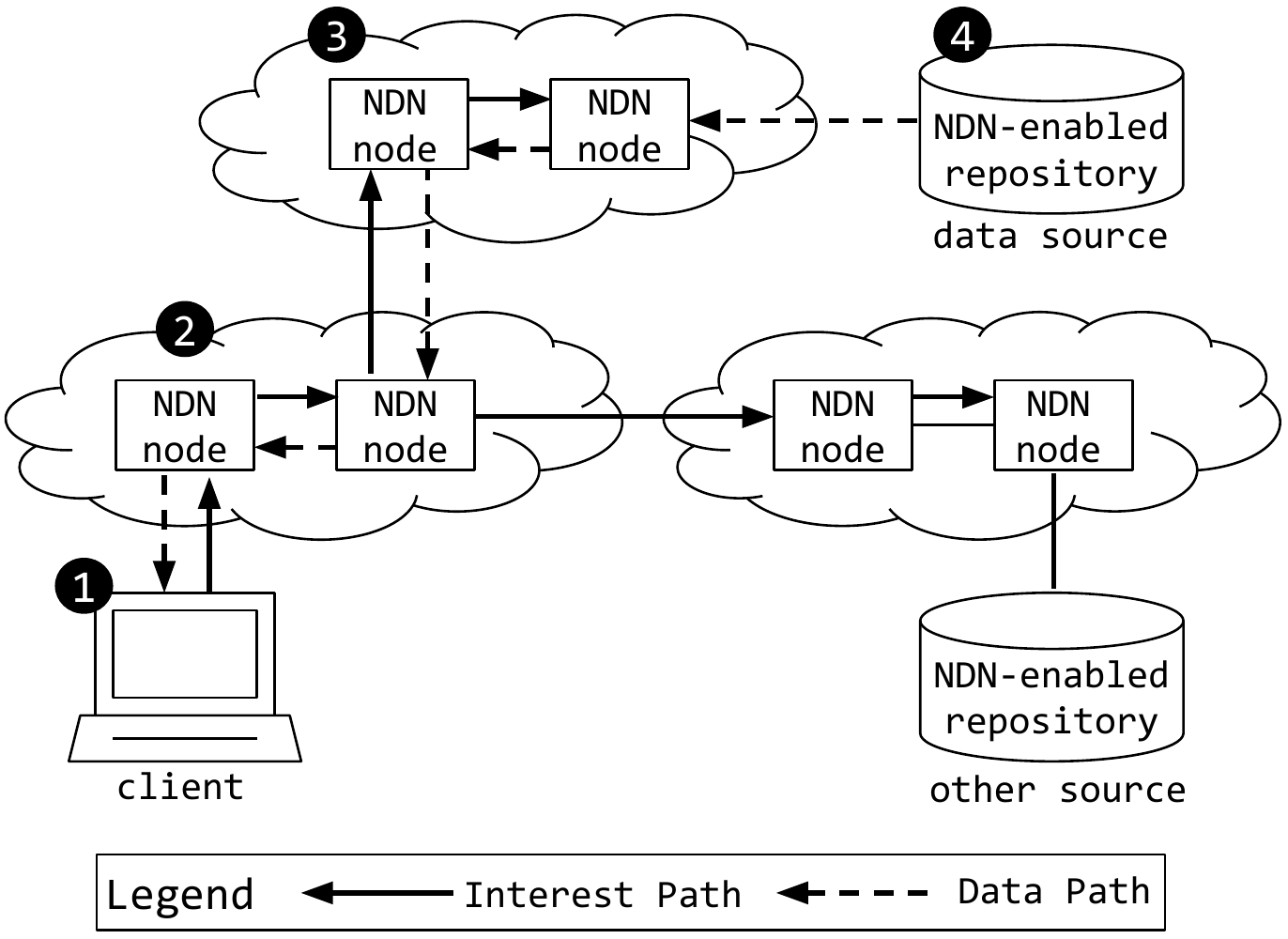}
\caption{Accessing data using data names in \ac{NDN} networks.}
\label{fig:NDNNetwork}
\end{figure}

In contrast to BitTorrent, \acf{NDN} is a current research topic of location-independent data access using information-centric principles \cite{jacobson_networking_2009}.
\ac{NDN} is also featured in the location-independent \ac{PID} approach presented in this paper to support a next-generation Internet technology.
In \ac{NDN}, data sets are enumerated through \textit{Data Names} that form a hierarchical name space \cite{jacobson_networking_2009}.
The working principle of \ac{NDN} is shown in \mbox{Figure~\ref{fig:NDNNetwork}.}
To access data from a client (step~\blackone) in the \textit{Named Data} space, an interest data package is sent through the network.
Based on the data name driven routing principles, the \ac{NDN} network directs the interest through the network (step \mbox{\blacktwo~and~\blackthree).}
If a \ac{NDN} node is found that owns a named data set \mbox{(step~\blackfour),} a data package is sent back along the interest path to reach the node, which stated the data request.
Hence, \ac{NDN} abstract from the network location and the data source remains opaque \cite{jacobson_networking_2009}.

\section{Magnet \acs{URI} Scheme}
\label{sec:MagnetURIScheme}

For embedding location access information into \ac{PID}, the Magnet \ac{URI} scheme is used as transport container, which is a work-in-progress specification for \emph{Magnet Links} \cite{mohr_magnet_2002}.
Magnet Links can store extensive information on accessing resources in networks, like \ac{HTTP} download \acs{URL}s, mirror server information, or peer-to-peer access data.
By applying this principle, Magnet Links are usable for describing digital resources and its content.
But as a descriptive access format, Magnet Links need a storage medium to be present on.
This could be a Web site with HTML content, an E-Mail message content, or -- like in our use case -- a Persistent Identifier.
To provide persistent access to digital data encoded in Magnet Links for a time of years, or possibly decades, a medium is needed that provides these properties.
Hence, persistency is not archived by the Magnet Links but provided by the \ac{PID} System that ensures long living existence of access information through its infrastructure, policies and replication partnerships.
If Magnet Links are stored on \emph{perishable} media like Web sites, they do not provide any advantage over common \ac{URL} access information.
The conjunction with \ac{PID} provides the additional benefit of long-term access from data hosted at suddenly moving data.
Besides the encoding of access information, one design goal of Magnet Links is to integrate features made available by local utility programs seamlessly into the storage medium like a Web site, by following best practices among the \ac{IETF} specifications for \ac{URN} \cite{sollins_rfc_1994}.
Although its lack of specification, Magnet Links are supported by numerous peer-to-peer tools and are the de facto standard in large file sharing communities, such as BitTorrent \cite{van_der_sar_download_2012}.
In BitTorrent-enabled Magnet Links, access information for downloading files from a decentralized peer-to-peer infrastructure are stored together with optional metadata and suggested file names \mbox{(c.f. Tab.~\ref{tab:MagnetURIScheme})}.
To initiate a download with a Magnet Link from a Web browser, a pseudo-protocol handler for the Magnet Link \ac{URL} format
\texttt{magnet:?xt=urn:<System>:<Access Information>}
is registered.
It passes the information to a location independent download client.

\begin{table}[ht]
\centering
\captionsetup{justification=centering}
\caption{Magnet URI scheme keys}
\begin{tabularx}{0.474\textwidth}{|l|l|X|}
\hline  \textbf{Key} & \textbf{Name} & \textbf{Purpose} \\ \hline
\texttt{as} & Acceptable Source & location dependent download URL\\ \hline
\texttt{dn} & Display Name      & file name \\ \hline
\texttt{kt} & Keyword Topic     & search key word \\ \hline
\texttt{tr} & Address Tracker   & optional tracker information for BitTorrent \\ \hline
\texttt{xl} & Exact Length      & size in bytes \\ \hline
\texttt{xt} & Exact Topic       & location independent access information in URN-format \\ \hline
\end{tabularx}
\label{tab:MagnetURIScheme}
\end{table}

\section{Location-Independent Data Access through PID}
\label{sec:LocationIndependentDataAccessThroughPID}

For creating a persistent resolution target in a Handle \ac{PID} that is based on the content of linked data set and not the network location like the target URL, we leverage the principles of location-independent data access with Magnet \acs{URI}-encoded access information.
In the first step in \mbox{Subsection~\ref{subsec:MagnetURISchemeExtentionForNDN},} we extend the Magnet \acs{URI} scheme into the domain of \ac{NDN} applications to support this cutting-edge data access technology.
By this, we can encode all access information for BitTorrent and \ac{NDN} as well as all systems listed in \mbox{Table~\ref{tab:MagnetLinkSchemeAdopters}} into one uniform scheme.

Then, in the second step in \mbox{Subsection~\ref{subsec:DataAccessThroughPIDWithPersistentResolutionTargets},} we propose an approach of embedding the Magnet Link \ac{URI} scheme into \acs{PID}s of an unmodified Handle System.
Starting from \mbox{Subsection~\ref{subsec:DataAccessUsingWebbrowser}} onward, we examine the \ac{PID} resolution of \acs{PID}s with enabled location-independent access data.

\subsection{Magnet \acs{URI} Scheme Extension for NDN}
\label{subsec:MagnetURISchemeExtentionForNDN}

Besides the already established usage of the Magnet URI scheme in peer-to-peer overlay networks that particularly allow BitTorrent location-independent data access, a next-generation access technology for supporting location-independent data access is integrated in our approach.
We extend the Magnet URI schema into the domain of content-centric networks, enabling support for Named Data Network data access through Magnet Links.
Thus, the Magnet URI schema is extended to store a \ac{NDN} data name that identifies a digital object within a \ac{NDN} network.
With the data name, the \ac{NDN} network can transport data from a source node holding the data back to the client node requesting the information through an interest \cite{jacobson_networking_2009}.
The location of the data is not important, as long as it is attached to the network through a reachable \ac{NDN} node.
The data name is encoded through an extended \texttt{xt} key that holds besides the data name also a checksum of the data name to detect data corruption.
The schema is \texttt{uid:ndn<DATANAME>.<CHECKSUM>}.
Unlike current host-based networks that use a \ac{PKI} to verify host identities through SSL certificates, \ac{NDN} data cannot verified through a trustful data location.
As a result, the Magnet Link also has to include the verification information needed to assure that the received content \emph{is} the requested content.
This is done by adding a cryptographic signature of the \ac{NDN} content in a separate \texttt{<SIGNATURE>} field, which is part of the second \texttt{xt} key extension.
Thus, verification needs to be done on content level using a public \ac{PKI}, which is in the scope of current \ac{NDN} research.
To get the certificate needed to verify the data, the \ac{NDN} access information for the certificate need to be added to the Magnet Link, too.
To obtain the certificate with the public key, we propose the \texttt{xt} key \texttt{uid:ndnsec<SIGNATURE>.<CERT\_DATANAME>} that allows the download of information for content verification.
By this, access \ac{NDN} access information can be encoded into a Magnet Link and also the genuineness of the obtained data can be verified through security information embedded into the Magnet Link.
The extension we provide for the Magnet URI scheme supporting \ac{NDN} is depicted in bold letters in Table~\ref{tab:MagnetLinkSchemeAdopters}.

\subsection{Embedding Magnet Links in Handle \acs{PID}s}

For this integration of Magnet Links into the Handle PID System maximum compatibility is paramount, as data dissemination has a very slow change momentum, owed to billions of PID-tagged data sets.
Hence, the usage of Magnet Links for location independent data access and its impact on the adaption in the Handle System is investigated.
By design, Handle supports hierarchical data types, identified by \mbox{UTF-8} named fields.
The data itself is organized as indexed, typed key-value pairs that store sequences of octets, which are preceded by its length in a 4-byte unsigned integer.
Like other \ac{PID} systems, the Handle System provides a \ac{URL} data type \newline\mbox{\texttt{0.TYPE/URL}} \cite{sun_rfc_2003}.
Supplemental services such as \ac{PID} \ac{HTTP} resolvers, also known as Handle proxies, use the \ac{URL} semantic to resolve \ac{PID}s into target URLs by using HTTP-forwarding with \ac{HTTP} status code \texttt{303}.

\begin{table}[ht]
\centering
\captionsetup{justification=centering}
\caption{Magnet Links Scheme Adopters (proposed schemes for Named Data Networking in bold font)}
\begin{tabularx}{0.474\textwidth}{|l|c|X|}
\hline  \textbf{System} & \textbf{URN} & \textbf{Value} \\ \hline
Gnutella2        & \texttt{sha1}         & file hash (SHA-1) \\ \hline
BitTorrent       & \texttt{btih}         & unique file identifier \\ \hline
Gnutella2        & \texttt{tiger}   & file hash (Tiger Tree Hash)\\ \hline
Kazaa            & \texttt{kzhash}       & file hash (proprietary) \\ \hline
\textbf{NDN Access}        & \texttt{\bfseries{ndn}}  & \textbf{DataName and Checksum (SHA256)} \\ \hline
\textbf{NDN Verification} & \texttt{\bfseries{ndnsec}} & \textbf{content signature \& public key data name (NDN specs.)} \\ \hline
\end{tabularx}
\label{tab:MagnetLinkSchemeAdopters}
\end{table}

Unlike \ac{URL}s, Magnet Links do not specify the data location but can be considered as \ac{URN}-like data classification.
Hence, the Handle \ac{PID} data type \mbox{\texttt{0.TYPE/URL}} does not fit semantically for Magnet Links, because \ac{URL} is a subset of \ac{URI} \cite{berners-lee_rfc_2005}.
As a result, an own data type \mbox{\texttt{MAGNET}} needs to be registered at a Handle \ac{PID} server that should be capable of holding Magnet Links.
To retrieve data from the \ac{PID} via location-independent technology, a Magnet Link can be placed into the Handle.
As Magnet Links fit into the \mbox{UTF-8} encoding of Handle values, they can be placed without any further encoding.
For Instance, a valid \mbox{\ac{DHT}-enabled} Magnet Link containing BitTorrent information for retrieving a file can be generated and stored in the \texttt{MAGNET} field of a Handle \ac{PID}.
Additionally, a Magnet Link-wrapped NDN Data Name can be placed into the \texttt{MAGNET} using \ac{URL} escaping according to \ac{NDN} name specifications \cite{yu_ndn_2014}.

\subsection{Data Access Through PID with Persistent Resolution Targets}
\label{subsec:DataAccessThroughPIDWithPersistentResolutionTargets}

\begin{figure}[ht]
\centering
\includegraphics[width=0.3\textwidth]{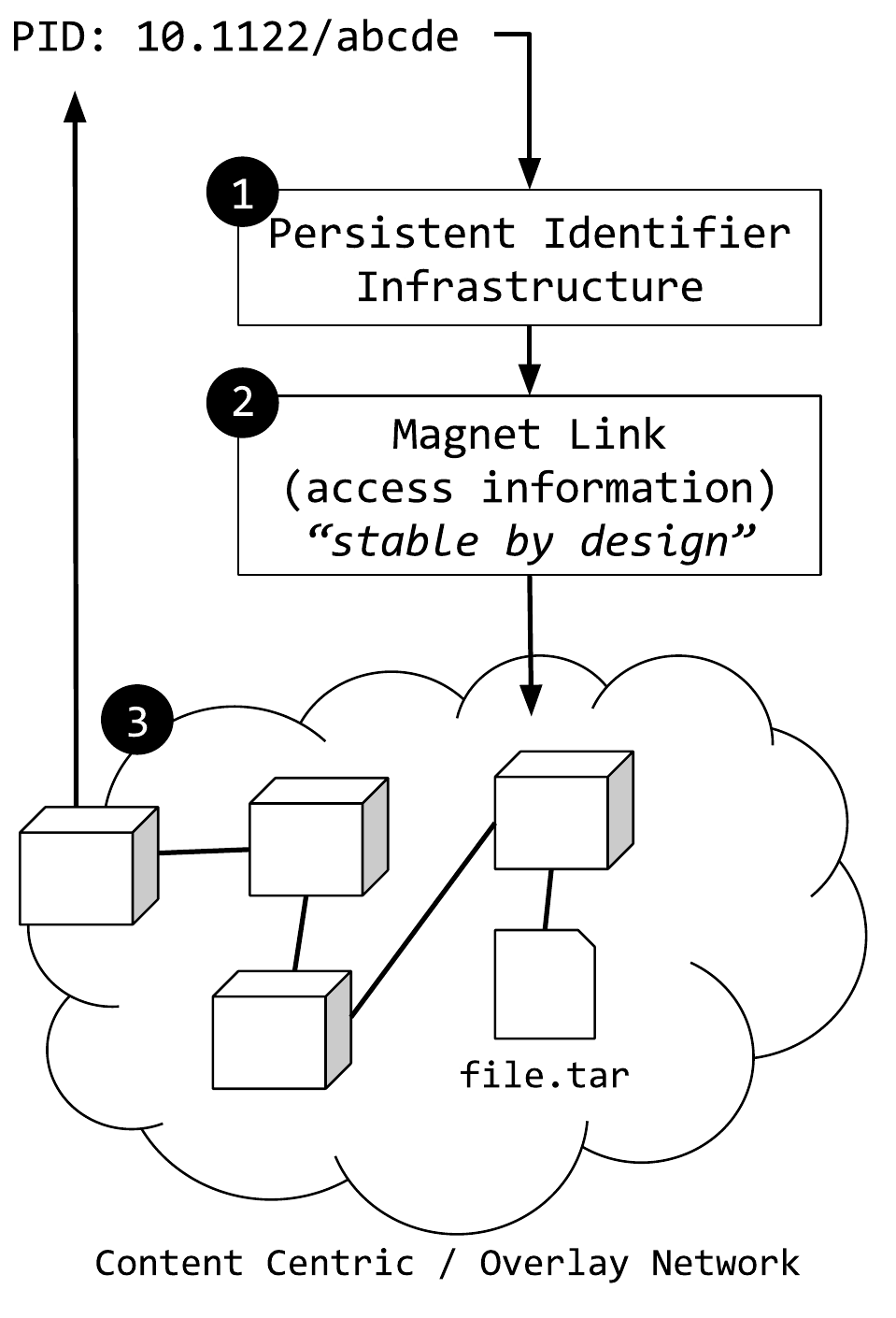}
\caption{Location independent access through PID relies on stable content-based access information.}
\label{fig:LIDABasedDataAcquisition}
\end{figure}

Handle PIDs that are equipped with a Magnet Link can be resolved like any other \acs{PID}s using the native Handle protocol.
This can be done either by using that protocol on top of \ac{TCP} or \ac{UDP}, or via a \acs{HTTP}-based proxy that answers resolution requests as a Web service.
The \emph{resolving} process for \acs{PID}s with persistent resolution targets works similar to the resolution process of location-based \acs{PID}s regarding the initial steps done within the Handle infrastructure.
Hence, Figures~\ref{fig:PIDMaintenance} and~\ref{fig:LIDABasedDataAcquisition} share the first initial \mbox{step~\blackone,} where the \ac{PID} is resolved by the Handle infrastructure.
In this step the \ac{GHR} determines the \ac{LHS}, which is responsible for a specific local sub-namespace (Handle prefix).
Then, the \ac{LHS} looks up the requested \ac{PID} in its database and returns the requested values to the client.
For resolving using location-based data access, the value with the type \mbox{\texttt{0.TYPE/URL}} is returned from the database and for resolving a with a persistent target location independent data access information in form of \mbox{\texttt{MAGNET}} is returned.

The new data access chain depicted in \mbox{Figure~\ref{fig:LIDABasedDataAcquisition}} is different from the location-based data access using the target URL shown in \mbox{Figure~\ref{fig:PIDMaintenance}.}
With the Magnet Link of the \ac{PID} acquired through the resolution process in \mbox{step~\blacktwo}~the data access is now handled by the overlay network in BitTorrent, or by the \ac{NDN} network \mbox{(step~\blackthree).}
The \ac{PID} resolving process is then a single redirection that leads to a starting download right after resolving, instead of multiple redirections using an entire chain of services for data access.
These connections rely on peer connections based on IP-addresses for BitTorrent and node connections for \ac{NDN}.
Thus, the number of steps is reduced to three; also fewer layers and infrastructure are required for accessing the data.
No central infrastructure is involved and the entire chain of location-based infrastructure is not needed anymore.
The only requirements for data access is a running \ac{PID} infrastructure and employing BitTorrent or \ac{NDN} software for sharing the data online.

\subsection{Creating and Resolving PIDs with Persistent Resolution Targets in Web Environments}

For a convenient and smooth resolution of \acs{PID}s in the context of the World Wide Web, querying and resolving of Handle \acs{PID}s is realized by using a proxy service that accepts requests in HTTP(S) and resolves and maintains \ac{PID} with the native protocol \cite{sun_rfc_2003}.
The Handle System maintainer, the \ac{CNRI}, provides a Handle Proxy Servlet that offers HTTP-based resolving.
The official Handle Proxy needs a small extension to resolve PIDs smoothly with HTTP into location-independent access information.
This is done by changing the resolving mechanism from \mbox{\texttt{0.TYPE/URL}} to \mbox{\texttt{MAGNET}} for the default value.
However, non-modified Handle Proxies also work, when explicitly querying for \mbox{\texttt{MAGNET}} or using URL rewriting appending query parameters on the \ac{HTTP} request.

For offering a Web-based creation of \acs{PID}s that features a Magnet Link as persistent resolution target, a Web service is a reasonable choice.
For every system that is offered for location-independent access technology, the Web service consumes either the native access information and encodes them into a Magnet Link or directly Magnet Links.
Then, the Web service is registering a Handle \ac{PID} at the \ac{LHS} using the native Handle protocol and updates the Handle with a \mbox{\texttt{MAGNET}} containing the Magnet Link.
In case of BitTorrent, the Web service consumes torrent files that contain along with the checksum and the name of the torrent also the infohash.
These information are parsed and embedded into the Magnet Links as persistent \ac{PID} resolution target using the current Magnet \ac{URI} scheme description.
For \ac{NDN} data access, the Web service consumes the data name and the SHA-256 checksum of the data set.
Furthermore, the Web service allows attaching \ac{NDN} verification information to \ac{PID} by extending the Magnet Link in the \ac{PID} containing the cryptographic signature, as well as the data name to obtain the public certificate of the owner.
The \ac{NDN} access information are encoded into Magnet Link using our proposed Magnet \ac{URI} scheme (c.f \mbox{Subsection~\ref{subsec:MagnetURISchemeExtentionForNDN})} in order to be embedded into the \ac{PID}.

\subsection{Data Access Through PID with Persistent Resolution Targets using a Web browser}
\label{subsec:DataAccessUsingWebbrowser}

Although, the Handle System can remain unmodified and the Handle proxy is only subject of optional minor modifications, the automatic resolution on HTTP clients provides solvable challenges.
To provide a smooth resolution, a \ac{HTTP} client like a Web browser needs to support multiple protocols through asking a protocol handler that determines the behaviour for resources outside \ac{HTTP} sphere, like \texttt{mailto:} resources that are forwarded to the E-Mail program.
For \texttt{magnet:} resources, the Magnet Link protocol handler is invoked~\cite{thaler_rfc_2015}.
Based on information stored in the Magnet Link value \texttt{\ac{xt}}, the Magnet Link Handler selects the appropriate application and passes all information necessary for the download.
Then the application is doing the heavy data download via the suggested access technology stored in the Magnet Link.

For invoking the process automatically in the Web browser, following steps are applied.
When resolving \ac{HTTP}-related \acs{URL}s, the proxy-based resolver responds with \ac{HTTP} status \texttt{"303}~\texttt{-~See Other"}.
According to the \ac{HTTP} standard RFC 7231, a \texttt{303} response to a \texttt{GET} request indicates that the server does have a representation of the target resource which can be transferred over \ac{HTTP} \cite{fielding_rfc_2014}.
In the case of normal location-based forwarding, the target \texttt{URL} field of the PID is resolved into a target server target URL, as the proxy server does not possess the data.
In the case of \acs{PID} holding location-independent access information, the \texttt{MAGNET} field is resolved into a forwarding to the overlay network of \ac{NDN} name space.
This \ac{HTTP}-forwarding indicates that the Handle proxy server does not posses the data and it is not reachable via \ac{HTTP}.
Hence, the use of Magnet \acs{URI}s in \acs{PID} for \ac{HTTP}-based resolution is in line with the HTTP standard and especially supports it with embedding additional information into the Magnet Link for a self descriptive data access.

\section{Implementation}
\label{sec:Implementation}

\subsection{Server Side}

For verifying our approach, we set up an entire stack of software component for verification.
On the server side, the stack consists of an unmodified Handle System (c.f.~\ref{subsubsec:LocalHandleSystem}) and a custom Web service called \emph{PID-Burner} (c.f.~\ref{subsubsec:PID-Burner}), which is able to create, update and resolve Magnet Link enabled \ac{PID}s based on our approach.

\subsubsection{Local Handle System}
\label{subsubsec:LocalHandleSystem}

The \ac{LHS} that hosts the Handle \acs{PID}s under a specific prefix does not require any modifications in the source code to run our approach for storing and serving \acs{PID}s with Magnet Links.
By default, the Handle System supports a list of preconfigured data types that are available as standard type set in every Handle System of a specific version.
The list of pre-configures data types contains types for realizing typical \ac{PID} scenarios with \texttt{EMAIL} and \texttt{URL} for location-based access.
But also special scenarios like target URL forwarding based on users' geolocations.
It should also contain a \texttt{URN} data type according to the Handle System documentation \cite{corporation_for_national_research_initiatives_4.9_2015}, but it -- unfortunately -- is not part of the preconfigured data types in the most recent version 8.1.0 \cite{corporation_for_national_research_initiatives_handle.net_2016}.

As none of these preconfigured data types are suitable or working for storing Magnet Links, we use the well-designed extensible type system of the Handle System architecture to register a new data type \texttt{MAGNET}.
By this, we only add a configuration item to the \ac{LHS} that has no impact on the existing type system and runs on Handle legacy systems, too.

\subsubsection{PID-Burner - Creating, Maintaining and Resolving of PIDs with Persistent Resolution Targets}
\label{subsubsec:PID-Burner}

The PID-Burner Web service allows creating, updating and resolving \acs{PID}s with Magnet Links embedded as persistent resolution targets.
It is implemented from scratch, but incorporates libraries and frameworks for \ac{PID} management, as well as BitTorrent libraries and initially created libraries for Handling \ac{NDN} access information and processing Magnet Links with the extensions proposed.
The PID-Burner engine is implemented in Python using the \emph{Bottle Web framework} from creating a \ac{REST} interface \cite{hellkamp_bottle:_2016}.
Besides the \ac{REST} interface it offers a JavaScript-based user interface for the Web browser.
As back-end for interacting with the Handle \ac{PID} service the EPIC-API v2 Web service from \ac{EPIC} is incorporated to create and update \acs{PID}s \cite{european_persistent_identifier_consortium_pidconsortium/epic-api-v2_2016}.
For processing BitTorrent access information contained in torrent files, \emph{libtorrent} Python bindings are used for extracting the necessary access information like the infohash, the file name and checksum \cite{norberg_libtorrent_2015}.
\ac{NDN} access information processing is done with a custom library, as well as the generation of Magnet Links.

For creating and updating \acs{PID}s with Magnet Links, the user can upload a torrent file that contains the BitTorrent access information.
These torrent files can be created from original files in BitTorrent programs like Transmission \cite{transmission_project_transmission_2016}.
\ac{NDN} access information are uploaded as \ac{JSON} data structures and can store the data name and the checksum.
The \ac{NDN} data names has to be determined by the user depending on its \ac{NDN} network topology.
The checksum can be computed using any checksumming tool like OpenSSL.
\ac{JSON} is used for \ac{NDN} access information encoding due to the lack of standardized access container formats in \ac{NDN} that are comparable to torrent container files.
The optional cryptographic verification information are attachable to the \ac{PID} using \ac{NDN} access information containing the cryptographic signature of the data and the \ac{NDN} data name to retrieve the X.509 certificate of the data signer.

For resolving Magnet Links enabled \acs{PID}s with \ac{HTTP}, the Web service implements a resolution functionality that is almost identical to the original Handle \ac{HTTP} proxy by \ac{CNRI} \cite{corporation_for_national_research_initiatives_handle.net_2015}.
As described in \mbox{Subsection~\ref{subsec:DataAccessThroughPIDWithPersistentResolutionTargets}}, the resolution process does not depend on target URLs, but rather uses the \ac{PID} value stored in the \texttt{MAGENT} data field of the \ac{PID}.
Hence, if a Web client asks the PID-Burner service for resolving, the Magnet Link with the access information is returned as HTTP status \texttt{303} \texttt{-} \texttt{See Other} for existing \acs{PID}s.
If the \ac{PID} does not contain a Magnet Link, the resolution is done against the \ac{URL} value of the \ac{PID} and PID-Burner behaves identical to the Handle HTTP proxy service.
The behaviour allows a maximum on compatibility towards the original Handle System also on \ac{HTTP} resolution.

\subsection{Client Side}

End-users who are interested in using Magnet Link-enabled \acs{PID}s for data access need client software that is able to process access information for location-independent access.
For BitTorrent-based access a client software is needed that is able to process Magnet Links.
To simulate end-user environment we are using a Gnome 3.16.2 Desktop on Fedora 22 together with Chromium 47.0.2526.106 as Web browser running on Intel i5-2400 with 8GB RAM.
As BitTorrent Software, we use an unmodified version of Transmission 2.92 \cite{transmission_project_transmission_2016} and use public BitTorrent infrastructure available to every Internet user (\ac{DHT} bootstrap servers) for file download.

For \ac{NDN} access, we provide an own Magnet Link adapter that process the \ac{NDN} access information and passes them to the \ac{NDN} download tools.
This \ac{NDN} Magnet Link adapter has been implemented in Python and is necessary to parse \ac{NDN} Magnet Links the based on our proposes schema extension.
For \ac{NDN} data hosting and downloading, we use the experimental \ac{NDN} Repo NG tool set that consists of \ac{NDN} server and download applications for exchanging data over a \ac{NDN} network~\cite{afanasyev_repo-ng:_2015}.
The Repo NG tool set is running in a private testbed at \ac{GWDG} that consists of six \ac{NDN} nodes.

\section{Evaluation and Discussion}
\label{sec:EvaluationAndDiscussion}

All approaches related to \ac{PID} systems have to provide interoperability at its highest degree to comply with the slow changing momentum of the Handle Infrastructure.

First, the Handle System is a very large distributed infrastructure with shared responsibilities.
The services consist of 1\,000 servers in 75 countries, which are operated by hundreds of organisations.
It currently holds over $>$100 million \acs{PID}s, owned by over 12 thousand registrants in 2015 \cite{corporation_for_national_research_initiatives_hdl_2015} \cite{international_doi_foundation_doi_2014}.
Hence, approaches that demand a fundamental change in the system have the challenges of convincing a large community.
In contrast, our approach presented in the paper operates on-top of the infrastructure and has no impact on existing \ac{PID} infrastructure.
If \ac{LHS} operators are interested in implementing \ac{PID}s with persistent resolution targets they just have to add a service that is able to resolve, maintain and update \texttt{Magnet} Links within \ac{PID}s as described in \ref{subsubsec:PID-Burner}.

Second, for data repository owners that use \ac{PID} for registering their data and \ac{HTTP}-based file distribution, our approach opens up new perspectives on data dissemination.
With our approach they can combine the advantages of location-independent access, peer-to-peer networks and \ac{PID} into a single concept.
For this, they have to create access information for their existing files and provide a location-independent upload point such as a BitTorrent Upload server.

Although our concept offers many advantages, we have to investigate its performance.
The evaluation has to be split into two parts; the first is the evaluation of the \ac{PID} access.
This is done by comparing the distribution of string lengths in PID target URLs against existing Magnet Link resources, checking whether Magnet Links will increase the size of PIDs.
If Magnet Links increase the size, resolution performance will decrease.
This can be explained with higher data transmission volumes and larger data sets that are to be handled by software stacks.

\begin{figure}[t]
\centering
\includegraphics[width=0.485\textwidth]{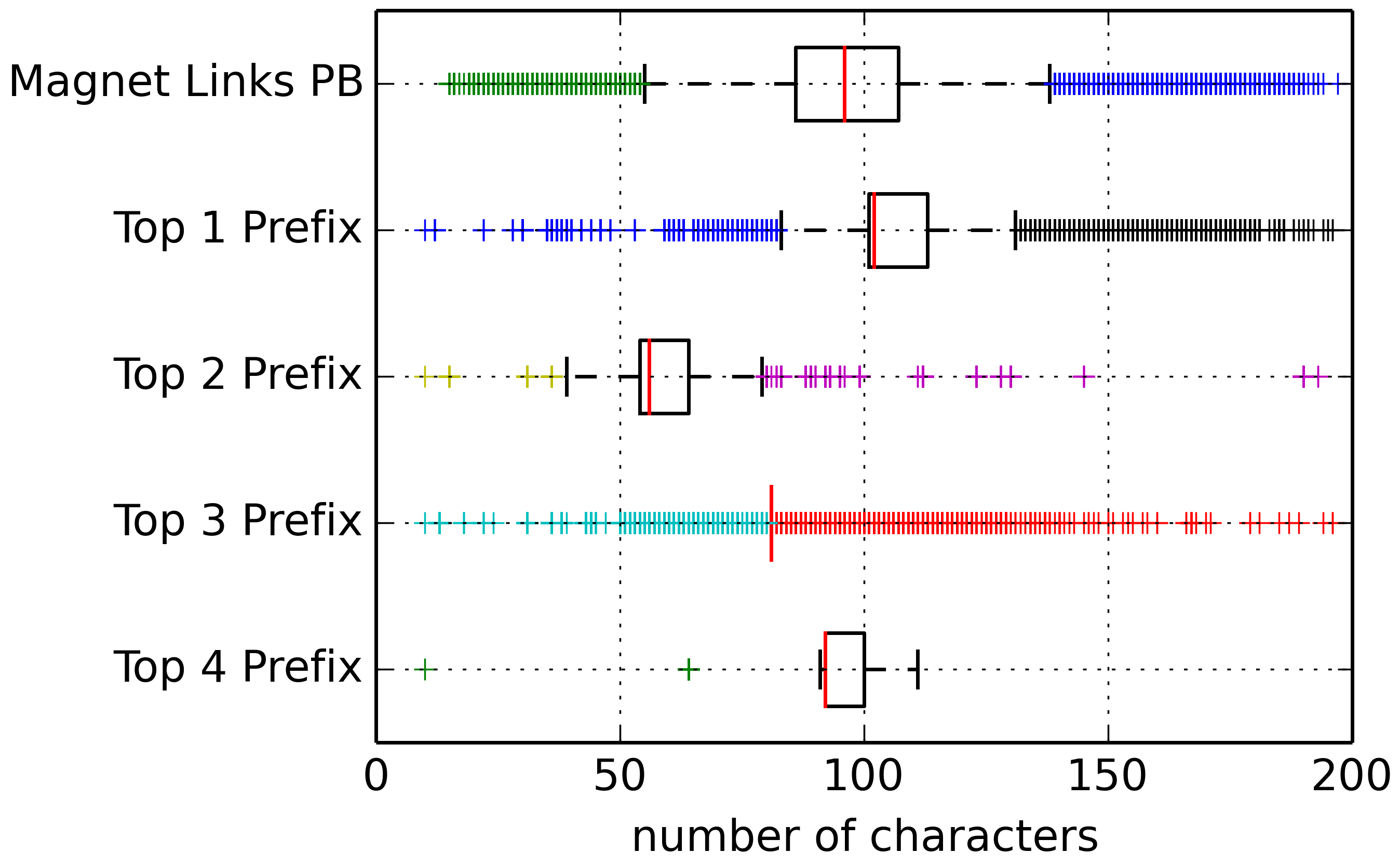}
\caption{Distribution of string length for Pirate Bay Magnet Links and PID target URLs for frequently resolved Handle prefixes (95\% data shown in plot, x-axis limited at 200 chars)}
\label{fig:CharacterDistribution}
\end{figure}

In Figure~\ref{fig:CharacterDistribution}, the string length distribution for BitTorrent Magnet URLs aggregated from the \textit{Pirate Bay} Web site is plotted as box-plot in row one.
The Pirate Bay data set (row one) is chosen as a benchmark data set, as it forms one of the largest data accumulation for heterogeneous files exchanged by BitTorrent.
For the string length analysis of the Pirate Bay data sets the tracker information have to be removed in order to provide clean analysis;
furthermore, the tracker information are not necessary to perform a download using \ac{DHT} techniques in BitTorrent.
It consists of 1\,643\,194 Magnet Links in total that consist of BitTorrent access information in the form of infohashes and file names \cite{van_der_sar_download_2012}.

In comparison to the access information in Magnet Links, we now compare the string lengths of \acs{PID}s (row two to five).
For this we use real-world data from network users who resolved \ac{PID}s at Handle Servers hosted at \ac{GWDG}.
The data set for \ac{PID} resolving data consists of 1.294.668 target URLs in total for a time span between June 2014 and August 2014.
The five Handle prefixes with the top-most resolution at the time span have been selected.

\begin{figure}[t]
\centering
\includegraphics[width=0.485\textwidth]{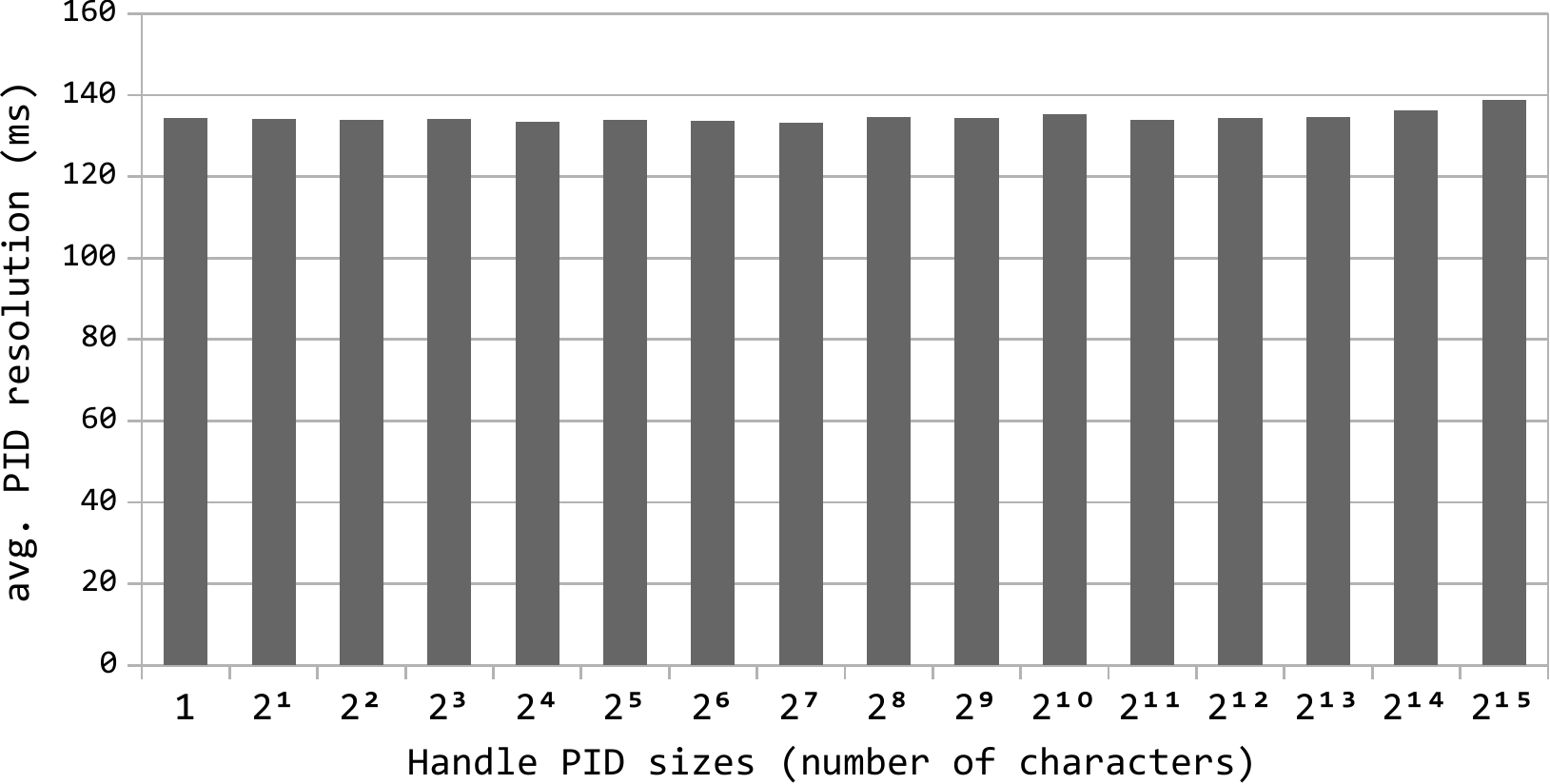}
\caption{Average resolution time of Handle PIDs with different target URL lengths measured at the GWDG \ac{LHS} for the Handle Prefix 11022.}
\label{fig:PIDvsResolution}
\end{figure}

The box plots in Figure~\ref{fig:CharacterDistribution} show that real-world Magnet Link collections share a comparable string length distribution with Handle PID target URLs.
To investigate the relation between \ac{PID} size and the \ac{PID} resolution performance, we measured the resolutions performance with \acs{PID}s of a defined size (c.f. \mbox{Figure~\ref{fig:PIDvsResolution}).}
The measurements were done at the \ac{LHS}, hosting the Handle Prefix 11022 at GWDG with suppressed caching support to measure raw resolution times.
It can be observed that the resolution time is slightly increasing for a number of milliseconds for extreme PID sizes with 32\,768 characters.
With the average \ac{PID} size derived from the Pirate Bay data set of 97 characters (c.f. with the $2^{7}$ bar in \mbox{Figure~\ref{fig:PIDvsResolution}),} the impact of Magnet Link usage in \ac{PID} is not perceivable to users.
As a result, the \ac{PID} replication and resolution can be expected similar to the existing target URL-based approach.
Embedding Magnet Links in \acs{PID}s has no significant impact on the size of \acs{PID}s and underlines the practical usability of the concepts presented in this paper.
Hence, Magnet Link enabled \acs{PID}s will not resolved into location-independent targets slower than current state-of-the-art Handle PIDs.
The \ac{PID} resolution time $t_{r}$ can be assumed to be identical.

The second part of the evaluation is the data download speed provided from the location-independent data access.
For BitTorrent and \ac{NDN} the time for bootstrapping $t_{b}$ the node and initiate a data download from the swarm is time intensive.
Bootstrapping includes in the case of BitTorrent joining the \ac{DHT}.
The latter is very fast, although state-of-the-art \ac{DHT} joining is accelerated through hard-coded bootstrapping servers.
For \ac{NDN} bootstrapping includes learning the node environments and the name routes.
Different \ac{NDN} bootstrapping algorithms are subject of current research \cite{hoque_nlsr:_2013}.

After successful bootstrapping the advantages can speed up data transfer.
If a peer is found that offer data access the download speed is at least comparable to existing location-based access that uses HTTP for download.
If more than one peer is found, the bandwidth can be utilized to provide simultaneous data transfers, too.
In this case the data volume $v$ is split into $n$ parts and the longest transfer $max()$ time of chunk is setting the overall transmission time as most time consuming partial transfer.
Hence, the transfer duration $d_{tr}$ of the first transfer through \ac{PID} can be estimated as
\begin{equation}
d_{tr} = t_{r} + t_{b} + max\left(\frac{v_{n}}{v/sec}\right)
\end{equation}
This parallelization results in higher download bandwidth and short transmission duration.
Thus, for large download volumes the data access starting from PID resolution to download completion is faster.
For small download volumes the completion time is longer in comparison to traditional data access through PID.
Traditional location-based approaches using HTTP with no multi-source download capabilities have a transfer duration of
\begin{equation}
d_{tr} = t_{r} + \sum_{n=0}^N \frac{v_{n}}{v/sec}
\end{equation}
where the duration is the sum of all partial volumes that are downloaded in serial.

Besides the performance aspects, it is observable that the \ac{PID} usage of the \ac{PID} infrastructure usage has an impact on the string length distribution of the target URLs.
The top 4 prefix has a dense character distribution that is caused by the fixed pattern of the target URL, where only parts of the URL are varying.
This is caused by the repository software that is managing the \acs{PID}s and uses IDs with similar length for ever data set.
In contrast, the top four prefix shows a sparse distribution of target URL string length caused by almost non-systematic target \ac{PID}s.
This distribution can be found by \ac{PID} System operator that offer \ac{PID} services to a large group of institutions like \ac{EPIC} \cite{european_persistent_identifier_consortium_pidconsortium/epic-api-v2_2016}.

\section{Future Work}

Despite our contribution to the \ac{PID} efforts in the Handle System challenges provide open research questions for future work.
The support for non-URL targets in HTTP-based Handle PID resolution could be moved into the existing Handle stack.
By this, resources can be directly linked in the typical workflows that end-user facilitate in their Web browsers.
Thus, location-independent access through \ac{PID}s in the Handle system could work with click, as simple as opening a PDF file.
Fortunately, the Handle \ac{PID} is very well designed and implemented by \ac{CNRI}, and the source code is available.

A number of challenges arises from the usage of Magnet Links.
The Magnet \ac{URI} scheme originates in the file sharing community and has evolved in the past decade.
Despite being a community effort, it is widely used by the most frequented file sharing search engines.
For usage in the research data community, Magnet \ac{URI} community contributions and drafts have to be collected and a standardisation effort needs to be started, \eg as an initial \ac{IETF} draft.
The relation between Magnet \ac{URI} and \ac{URN} will facilitate the standardisation efforts and help to improve the reputation of Magnet Links.

Moreover, we propose using Magnet Links in the \ac{NDN} community for encoding of full access information.
Magnet Links could be one appropriate container format for transmitting \ac{NDN} access information, which is closer to the original idea of object identification, based on \ac{URN}, rather than location identification done with current \ac{NDN} concepts based on \ac{URL}.
Similar approaches have been provided for other content-centric network like Open NetINF and proposed at IETF \cite{farrell_rfc_2014}.

\section{Conclusion}
\label{sec:Conclusion}

The integration of location independent data access in persistent identifiers is feasible without major modification of PID infrastructure.
The Magnet URI scheme is suitable container for storing application independent access information inside Handle PIDs although it currently lacks IETF standardisation.
As illustrated in our evaluation, their usage has no major implication on PID System operation and usage.
Employing Magnet Link enables the creation of maintenance free PIDs, which do not require target URL adjustments and thus reduce the residual efforts on data repository owner sides.
With the support of overlay network usage and NDN data access, \emph{better} data dissemination is achievable with augmented resilience through multi-peer data hosting.

\section*{Acknowledgments}
We like to thank Sven Bingert from the \acl{EPIC} for his support on the Handle evaluation infrastructure used for this paper.
We acknowledge research funding by Deutsche Forschungsgemeinschaft (DFG) under grant SFB 963/2 ``Astrophysical Flow Instabilities and Turbulence'', projects INF.

\raggedright
\bibliographystyle{myIEEEtran}
\bibliography{Literature}

\begin{thebibliography}{10}
\providecommand{\url}[1]{#1}
\csname url@samestyle\endcsname
\providecommand{\newblock}{\relax}
\providecommand{\bibinfo}[2]{#2}
\providecommand{\BIBentrySTDinterwordspacing}{\spaceskip=0pt\relax}
\providecommand{\BIBentryALTinterwordstretchfactor}{4}
\providecommand{\BIBentryALTinterwordspacing}{\spaceskip=\fontdimen2\font plus
\BIBentryALTinterwordstretchfactor\fontdimen3\font minus
  \fontdimen4\font\relax}
\providecommand{\BIBforeignlanguage}[2]{{%
\expandafter\ifx\csname l@#1\endcsname\relax
\typeout{** WARNING: IEEEtran.bst: No hyphenation pattern has been}%
\typeout{** loaded for the language `#1'. Using the pattern for}%
\typeout{** the default language instead.}%
\else
\language=\csname l@#1\endcsname
\fi
#2}}
\providecommand{\BIBdecl}{\relax}
\BIBdecl

\bibitem{paskin_digital_2011}
N.~Paskin, ``Digital {Object} {Identifier} ({DOI}) {System},'' in
  \emph{Encyclopedia of {Library} and {Information} {Sciences}}, 3rd~ed.\hskip
  1em plus 0.5em minus 0.4em\relax Boca Raton, FL: CRC Press, 2011, pp.
  1586--1592.

\bibitem{ahlgren_survey_2012}
B.~Ahlgren, C.~Dannewitz, C.~Imbrenda, D.~Kutscher, and B.~Ohlman, ``A survey
  of information-centric networking,'' \emph{IEEE Comm. Magazine}, vol.~50,
  no.~7, pp. 26--36, 2012. doi: 10.1109/MCOM.2012.6231276

\bibitem{van_der_sar_pirate_2009}
\BIBentryALTinterwordspacing
E.~Van~der Sar, ``The {Pirate} {Bay} {Tracker} {Shuts} {Down} for {Good},''
  Nov. 2009. [Online]. Available:
  \url{https://torrentfreak.com/the-pirate-bay-tracker-shuts-down-for-good-091117/}
\BIBentrySTDinterwordspacing

\bibitem{sollins_pervasive_2012}
K.~Sollins, ``Pervasive persistent identification for {Information} centric
  networking,'' in \emph{Proc. of the {Second} {Edition} of the {ICN}
  {Workshop} on {Information}-centric {Networking}}.\hskip 1em plus 0.5em minus
  0.4em\relax Helsinki, Finland: ACM, 2012. doi: 10.1145/2342488.2342490 pp.
  1--6.

\bibitem{koponen_data-oriented_2007}
T.~Koponen, M.~Chawla, B.-G. Chun, A.~Ermolinskiy, K.~H. Kim, S.~Shenker, and
  I.~Stoica, ``A {Data}-oriented (and {Beyond}) {Network} {Architecture},'' in
  \emph{Proceedings of the 2007 {Conference} on {Applications}, {Technologies},
  {Architectures}, and {Protocols} for {Computer} {Communications}}, ser.
  {SIGCOMM} '07.\hskip 1em plus 0.5em minus 0.4em\relax New York, NY, USA: ACM,
  2007. doi: 10.1145/1282380.1282402 pp. 181--192.

\bibitem{dannewitz_opennetinf_2012}
C.~Dannewitz, M.~Herlich, and H.~Karl, ``{OpenNetInf} - prototyping an
  information-centric {Network} {Architecture},'' in \emph{Proceedings of the
  37th {IEEE} {Conference} on {Local} {Computer} {Networks} {Workshops} 2012},
  Clearwater, USA, Oct. 2012. doi: 10.1109/LCNW.2012.6424044 pp. 1061--1069.

\bibitem{fotiou_illustrating_2012}
N.~Fotiou, D.~Trossen, and G.~C. Polyzos, ``Illustrating a publish-subscribe
  {Internet} architecture,'' \emph{Telecommunication Systems}, vol.~51, no.~4,
  pp. 233--245, Dec. 2012. doi: 10.1007/s11235-011-9432-5

\bibitem{dannewitz_secure_2010}
C.~Dannewitz, J.~Golic, B.~Ohlman, and B.~Ahlgren, ``Secure {Naming} for a
  {Network} of {Information},'' in \emph{Proc. of {IEEE} {Conference} on
  {Computer} {Communications} {INFOCOM}}.\hskip 1em plus 0.5em minus
  0.4em\relax San Diego, USA: IEEE, 2010. doi: 10.1109/INFCOMW.2010.5466661 pp.
  1--6.

\bibitem{haun_towards_2013}
S.~Haun and A.~Nürnberger, ``Towards {Persistent} {Identification} of
  {Resources} in {Personal} {Information} {Management},'' in \emph{Proc. of the
  3rd {International} {Workshop} on {Semantic} {Digital} {Archives} ({SDA}
  2013)}, vol. 1091.\hskip 1em plus 0.5em minus 0.4em\relax Valetta, Malta:
  CEUR Workshop Proc., Sep. 2013, pp. 73--80.

\bibitem{hilse_implementing_2006}
H.-W. Hilse and J.~Kothe, \emph{Implementing persistent identifiers: overview
  of concepts, guidelines and recommendations}.\hskip 1em plus 0.5em minus
  0.4em\relax London: CERL, 2006.

\bibitem{cruse_general_2016}
\BIBentryALTinterwordspacing
T.~Cruse, ``General {Assembly} 2016, moving {DataCite} forward,'' 2016.
  [Online]. Available: \url{https://blog.datacite.org/general-assembly-2016/}
\BIBentrySTDinterwordspacing

\bibitem{_datacite_2016}
\BIBentryALTinterwordspacing
``{DataCite} {Metadata} {Stats},'' Apr. 2016. [Online]. Available:
  \url{http://stats.datacite.org/}
\BIBentrySTDinterwordspacing

\bibitem{fenner_digging_2015}
\BIBentryALTinterwordspacing
M.~Fenner, ``Digging into {Metadata} using {R},'' Aug. 2015. [Online].
  Available: \url{https://blog.datacite.org/digging-into-data-using-r/}
\BIBentrySTDinterwordspacing

\bibitem{cohen_bittorrent_2013}
\BIBentryALTinterwordspacing
B.~Cohen, ``The {BitTorrent} {Protocol} {Specification} - {BEP} 3,'' Oct. 2013.
  [Online]. Available: \url{http://www.bittorrent.org/beps/bep\_0003.html}
\BIBentrySTDinterwordspacing

\bibitem{loewenstern_bittorrent_2013}
\BIBentryALTinterwordspacing
A.~Loewenstern and A.~Norberg, ``The {BitTorrent} {Protocol} {Specification} -
  {BEP} 5,'' Mar. 2013. [Online]. Available:
  \url{http://www.bittorrent.org/beps/bep\_0005.html}
\BIBentrySTDinterwordspacing

\bibitem{maymounkov_kademlia:_2002}
P.~Maymounkov and D.~Mazières, ``Kademlia: {A} {Peer}-to-{Peer} {Information}
  {System} {Based} on the {XOR} {Metric},'' in \emph{Peer-to-{Peer}
  {Systems}}.\hskip 1em plus 0.5em minus 0.4em\relax Berlin, Heidelberg:
  Springer, 2002, vol. 2429, pp. 53--65.

\bibitem{jacobson_networking_2009}
V.~Jacobson, D.~K. Smetters, J.~D. Thornton, M.~F. Plass, N.~H. Briggs, and
  R.~L. Braynard, ``Networking named content,'' in \emph{Proceedings of the 5th
  international conference on {Emerging} networking experiments and
  technologies}.\hskip 1em plus 0.5em minus 0.4em\relax Rome, Italy: ACM Press,
  Dec. 2009. doi: 10.1145/1658939.1658941 p.~1.

\bibitem{mohr_magnet_2002}
\BIBentryALTinterwordspacing
G.~Mohr, ``Magnet {URI} - {Draft} {Tech} {Overview}/{Spec},'' Jun. 2002.
  [Online]. Available:
  \url{http://magnet-uri.sourceforge.net/magnet-draft-overview.txt}
\BIBentrySTDinterwordspacing

\bibitem{sollins_rfc_1994}
\BIBentryALTinterwordspacing
K.~Sollins and L.~Masinter, ``{RFC} 1737 - {Functional} {Requirements} for
  {Uniform} {Resource} {Names},'' Dec. 1994. [Online]. Available:
  \url{https://tools.ietf.org/html/rfc1737}
\BIBentrySTDinterwordspacing

\bibitem{van_der_sar_download_2012}
\BIBentryALTinterwordspacing
E.~Van~der Sar, ``Download a {Copy} of {The} {Pirate} {Bay}, {It}'s {Only} 90
  {MB},'' Feb. 2012. [Online]. Available:
  \url{https://torrentfreak.com/download-a-copy-of-the-pirate-bay-its-only-90-mb-120209/}
\BIBentrySTDinterwordspacing

\bibitem{sun_rfc_2003}
\BIBentryALTinterwordspacing
S.~X. Sun, S.~Reilly, and B.~Boesch, ``{RFC} 3650 - {Handle} {System}
  {Overview},'' 2003. [Online]. Available:
  \url{https://tools.ietf.org/html/rfc3650}
\BIBentrySTDinterwordspacing

\bibitem{berners-lee_rfc_2005}
\BIBentryALTinterwordspacing
T.~Berners-Lee, R.~Fielding, and L.~Masinter, ``{RFC} 3986 - {Uniform}
  {Resource} {Identifier} ({URI}): {Generic} {Syntax},'' 2005. [Online].
  Available: \url{https://tools.ietf.org/html/rfc3986}
\BIBentrySTDinterwordspacing

\bibitem{yu_ndn_2014}
\BIBentryALTinterwordspacing
Y.~Yu, A.~Afanasyev, Z.~Zhu, and L.~Zhang, ``{NDN} {Technical} {Memo}: {Naming}
  {Conventions} - {NDN}, {Technical} {Report} {NDN}-0023, {Revision} 1,'' Jul.
  2014. [Online]. Available:
  \url{http://named-data.net/wp-content/uploads/2014/08/ndn-tr-22-ndn-memo-naming-conventions.pdf}
\BIBentrySTDinterwordspacing

\bibitem{thaler_rfc_2015}
\BIBentryALTinterwordspacing
D.~Thaler, T.~Hansen, and T.~Hardie, ``{RFC} 7595 - {Guidelines} and
  {Registration} {Procedures} for {URI} {Schemes},'' Jun. 2015. [Online].
  Available: \url{https://tools.ietf.org/html/rfc7595}
\BIBentrySTDinterwordspacing

\bibitem{fielding_rfc_2014}
\BIBentryALTinterwordspacing
R.~Fielding and J.~Reschke, ``{RFC} 7231 - {Hypertext} {Transfer} {Protocol}
  ({HTTP}/1.1): {Semantics} and {Content},'' Jun. 2014. [Online]. Available:
  \url{http://tools.ietf.org/html/rfc7231\#section-6.4.4}
\BIBentrySTDinterwordspacing

\bibitem{corporation_for_national_research_initiatives_4.9_2015}
\BIBentryALTinterwordspacing
{CNRI}, ``4.9 {Handle} {Value} {Line} {Format},'' in \emph{{HANDLE}.{NET}
  (version 8.1) {Technical} {Manual}}, Nov. 2015, pp. 28--29. [Online].
  Available: \url{https://hdl.handle.net/20.1000/105}
\BIBentrySTDinterwordspacing

\bibitem{corporation_for_national_research_initiatives_handle.net_2016}
\BIBentryALTinterwordspacing
------, ``{Handle.Net software (HN\_v8.1)},'' 2015. [Online]. Available:
  \url{http://www.handle.net/download\_hnr.html}
\BIBentrySTDinterwordspacing

\bibitem{hellkamp_bottle:_2016}
\BIBentryALTinterwordspacing
M.~Hellkamp, ``Bottle: {Python} {Web} {Framework},'' Feb. 2016. [Online].
  Available: \url{http://bottlepy.org/docs/0.12/}
\BIBentrySTDinterwordspacing

\bibitem{european_persistent_identifier_consortium_pidconsortium/epic-api-v2_2016}
\BIBentryALTinterwordspacing
{European Persistent Identifier Consortium}, ``pidconsortium/{EPIC}-{API}-v2,''
  Mar. 2016. [Online]. Available:
  \url{https://github.com/pidconsortium/EPIC-API-v2}
\BIBentrySTDinterwordspacing

\bibitem{norberg_libtorrent_2015}
\BIBentryALTinterwordspacing
A.~Norberg, ``libtorrent python binding,'' 2015. [Online]. Available:
  \url{http://www.rasterbar.com/products/libtorrent/python\_binding.html}
\BIBentrySTDinterwordspacing

\bibitem{transmission_project_transmission_2016}
\BIBentryALTinterwordspacing
{Transmission Project}, ``Transmission,'' Mar. 2016. [Online]. Available:
  \url{https://www.transmissionbt.com/}
\BIBentrySTDinterwordspacing

\bibitem{corporation_for_national_research_initiatives_handle.net_2015}
\BIBentryALTinterwordspacing
{CNRI}, ``Handle.{Net} {Registry},'' 2015. [Online]. Available:
  \url{https://www.handle.net/proxy\_servlet.html}
\BIBentrySTDinterwordspacing

\bibitem{afanasyev_repo-ng:_2015}
\BIBentryALTinterwordspacing
A.~Afanasyev, S.~Chen, W.~Shang, and J.~Shi, ``repo-ng: {Next} generation of
  {NDN} repository,'' Nov. 2015. [Online]. Available:
  \url{https://github.com/named-data/repo-ng}
\BIBentrySTDinterwordspacing

\bibitem{corporation_for_national_research_initiatives_hdl_2015}
\BIBentryALTinterwordspacing
{CNRI}, ``{HDL}® {Identifier} and {Resolution} {Services},'' Oct. 2015.
  [Online]. Available: \url{http://www.handle.net/factsheet.html}
\BIBentrySTDinterwordspacing

\bibitem{international_doi_foundation_doi_2014}
\BIBentryALTinterwordspacing
{International DOI Foundation}, ``{DOI} {News} - {September} 2014,'' Sep. 2014.
  [Online]. Available: \url{http://www.doi.org/news/DOI\_News\_Sep14.pdf}
\BIBentrySTDinterwordspacing

\bibitem{hoque_nlsr:_2013}
A.~K. M.~M. Hoque, S.~O. Amin, A.~Alyyan, B.~Zhang, L.~Zhang, and L.~Wang,
  ``{NLSR}: {Named}-data {Link} {State} {Routing} {Protocol},'' in \emph{Proc.
  of the 3rd {ACM} {SIGCOMM} {Workshop} on {Information}-centric {Networking}
  {ICN}}.\hskip 1em plus 0.5em minus 0.4em\relax New York, USA: ACM, 2013. doi:
  10.1145/2491224.2491231 pp. 15--20.

\bibitem{farrell_rfc_2014}
\BIBentryALTinterwordspacing
S.~Farrell, C.~Dannewitz, P.~Hallam-Baker, D.~Kutscher, and B.~Ohlman, ``{RFC}
  6920 - {Naming} {Things} with {Hashes},'' Apr. 2014. [Online]. Available:
  \url{https://tools.ietf.org/html/rfc6920}
\BIBentrySTDinterwordspacing

\end{thebibliography}

\end{document}